\documentstyle[aps,preprint]{revtex}

\tightenlines
\begin{document}

\title{Sandpiles 
with height restrictions}
			    
\author{Ronald Dickman$^{1,\dagger}$,
T\^ania Tom\'e$^1$, and M\'ario J. de Oliveira$^2$}
\address{
$^1$ Departamento de F\'{\i}sica, ICEx,
Universidade Federal de Minas Gerais,
Caixa Postal 702,
30161-970 Belo Horizonte, MG, Brazil\\
$^2$ Instituto de F\'{\i}sica,
Universidade de S\~ao Paulo,
Caixa Postal 66318,
05315-970 S\~ao Paulo, SP, Brazil\\
}
\date{\today}

\maketitle
\begin{abstract}
We study stochastic sandpile models with a height restriction in one
and two dimensions.
A site can topple if it has a height of two, as in Manna's model, 
but, in contrast to previously studied sandpiles, here the height 
(or number of particles per site),
cannot exceed two.  This yields a considerable simplification 
over the unrestricted case, in which the number of states
per site is unbounded.
Two toppling rules are considered: in one, the particles are 
redistributed independently, while the other involves some cooperativity.
We study the fixed-energy system (no input or loss of particles) 
using cluster approximations and extensive simulations, and find
that it exhibits a
continuous phase transition to an absorbing state at a critical value 
$\zeta_c$ of the particle density.  
The critical exponents agree with those of the
unrestricted Manna sandpile.
\end{abstract}



\date{\today}

\section{Introduction}

Sandpile models are the prime example of self-organized 
criticality (SOC) \cite{btw,dhar99}, 
a control mechanism that forces a system with an absorbing-state phase
transition to its critical point \cite{bjp,granada,cancun}, leading to
scale-invariance in the apparent absence of parameters \cite{ggrin}.  
SOC in a slowly-driven sandpile corresponds to an
absorbing-state phase transition in a model having the 
same local dynamics, but a fixed number of 
particles \cite{bjp,tb88,vz,dvz,vdmz}.  The latter class of models have
come to be called fixed-energy sandpiles (FES).  
While most studies of sandpiles have probed
the driven case \cite{dhar99,priez}, there is great interest in understanding 
the scaling properties of FES models as well \cite{dvz,mdvz,chessa,carlson}.  
In this paper we study FES with a height restriction.

From the theoretical standpoint, an inconvenient feature of sandpile 
models is that the number of particles per site is unbounded.  This 
complicates attempts to derive cluster 
approximations and continuum descriptions.  
In Manna's stochastic sandpile \cite{manna,manna2},
a site with $z \geq 2$ particles is active, i.e., can topple,
sending two particles to
neighboring sites.  This suggests restricting the number of particles
per site to $z =$ 0, 1 or 2.  In this work we study such a model, in one and
two dimensions, with the goal of establishing its critical properties.
Analyses of FES without a height restriction reveal that they
exhibit a phase transition between an absorbing and an active state
as the particle density $\zeta$ is increased beyond a critical value
\cite{bjp,fes2,m1d}; we find the same to be true of the restricted-height
models.  Thus the restricted model has nontrivial critical behavior, and
represents, due to its simplicity, an attractive system for further
theoretical analysis.  Moreover, a detailed study allows us to address
questions of universality in sandpiles, and, more 
generally, of absorbing-state phase transitions in systems with a conserved
density \cite{ale00}.
The balance of this paper is organized
as follows. 
In Sec. II we define the models, followed by a discussion of cluster
approximations in Sec. III.  
Numerical results are  
analyzed in Sec. IV, and in Sec. V we summarize 
our findings.

\section{Models}

The models are defined on a hypercubic
lattice with periodic boundaries: a ring of $L$ sites in one dimension, 
a square lattice of $L \times L$ sites in $2d$. The
configuration is specified by the number of particles
$z_i = 0, 1$, or 2 at each site $i$; sites with $z_i \!=\! 2$ are 
{\it active}, and have a toppling rate of unity.  The 
continuous-time (i.e., sequential), Markovian dynamics 
consists of a series of toppling events at individual sites.
(Maintaining the restiction $z \leq 2$ would be quite 
complicated in a simultaneous-update scheme.)
When site $i$ topples, two particles attempt to move to randomly 
chosen nearest neighbors $j$ and $j'$ of $i$.  
($j$ and $j'$ need not be distinct.)
Each particle transfer is
accepted so long as it does not lead to a site having 
more than two particles.  
The next site to topple is 
chosen at random from a list of active sites, which must naturally be 
updated following each event.  
The time increment associated with 
each toppling is $\Delta t = 1/N_A$, where $N_A$ is the number of active 
sites just prior to the event.
$\Delta t$ is the mean waiting time to the next event, if we were
to choose sites blindly, instead of using a list.
In this way, $N_A$ sites topple per unit time, consistent with each
active site having a unit rate of toppling.

We consider two stochastic toppling rules.  In one, 
the two particles released when a site topples move independently.
Any particle attempting to move to a site harboring two particles is
sent back to the toppling site.  (Thus an attempt to send two particles
from site $j$ to site $k$, with $z_k = 1$, results in $z_k\!=\!2$
and $z_j=1$.)   We study this {\it independent} toppling 
rule in both one and two dimensions.
In the other, {\it cooperative} rule, transitions that would 
transfer fewer than the maximum possible number of particles are avoided.  
The cooperative rule is studied in one dimension only.
Transition probabilities for the two rules are listed in Table I.

\section{Cluster approximations}

We have derived cluster approximations for the independent
toppling rule at the one-site (i.e., simple mean-field theory)
and two-site levels.
While the height restriction complicates the analysis of
transitions, it confers the advantage of a
strict limit on the the number
of variables.  (To study the unrestricted sandpile using cluster
approximations one
must impose a cutoff on the height distribution \cite{bjp}.)

\subsection{One-site approximation}

At this level of approximation there are
three variables, $p_n$, with $n = 0, 1$ or 2,
representing the probability of a site having exactly
$n$ particles.   It is convenient to use the shorthand
notation $p_n \equiv (n)$.
There is only one independent variable, due to the constraints
of normalization, $(0) + (1) + (2) = 1$, and of fixed density,
$\zeta = (1) + 2(2)$.

We begin the analysis by enumerating, in Fig. 1,
the possible transitions between states of a single site. 
Each transition requires a specific local configuration (of two or
three sites, depending on the process), and a particular redistribution
of the two particles liberated when the active site topples.  The
local configuration and the choice of redistribution
are independent events.  In the one-site approximation all joint 
probabilities for two or more sites are factorized:
$(ij) \to (i)(j)$ and $(ijk) \to (i)(j)(k)$.  

To illustrate how transition rates are evaluated we
consider some examples.  The transition
0 $\to $ 1 requires the initial configuration \fbox{0} \fbox{2}, i.e., an
empty site with an active neighbor.  Exactly one of the
two particles must jump to the empty site; in d dimensions this occurs
with probability $(2d\!-\!1)/2d^2$.  Thus the rate of transitions 0$\to $1
is 
\[
2d \frac{2d\!-\!1}{2d^2} (0)(2) = \frac{2d\!-\!1}{d} (0)(2)
\]
where the factor $2d$ represents the number of nearest neighbors.

Consider now the transition 2 $\to$ 1.  There are two mutually
exclusive paths by which it can be realized.  In one, both particles jump
to the same neighbor (the probability for this event is $1/4d^2$); 
if the neighbor bears a single particle, then only
one particle will be transferred, as required.  
Thus the initial configuration must
be \fbox{2} \fbox{1} and the rate for this path is
$ (2)(1)/2d$.
In the other path, the particles jump to distinct sites 
(the probability for this is $1/2d^2$),
one of which must already have two particles, while the other must have 
fewer than two.  The required initial configuration is therefore 
\fbox{2} \fbox{2} \fbox{$\not \!2$},
where $\not \!2$ denotes a site with $z<2$.  The rate for this path is
$(2\!-\!d^{-1}) (2)^2 (\not \!2)$. 
Evaluating the rates for the remaining transitions we obtain the
equations of motion:
\begin{equation}
\frac{d}{dt} (0) = \frac{2d\!-\!1}{2d} (2) 
[ (\not \!2)^2 -2(0)]   \;,
\end{equation}
\begin{equation}
 \;\;\;\;\; \frac{d}{dt} (1) = \frac{2d\!-\!1}{d} (2) [(0) + (2) (\not \!2) -(1)]
\end{equation}
and
\begin{equation}
\;\;\;\;\;\;\;\;\;\; \frac{d}{dt} (2) = 
\frac{2d\!-\!1}{2d} (2) [2(1) - (\not \!2)^2 - 2 (2) (\not \!2)]  \;.
\end{equation}
After eliminating the variables (0) and (1),
a simple calculation shows that the stationary density of
active sites is
\begin{equation}
(2) = 2 - \sqrt{5-2\zeta}
\end{equation}
which implies $\zeta_c = 1/2$ regardless of $d$.

\subsection{Two-site approximation}

The dynamical variables are now the nearest-neighbor (NN)
joint probabilities $(ij)$ with $i,j = 0$, 1, or 2.
There are four independent variables, due to the
symmetry $(ij) = (ji)$ (for $i \neq j$) and the two
relations noted previously.
The allowed transitions between configurations of a NN pair of sites are
shown in Fig. 2.

Consider, for example, the transition 00 $\to$ 01.  The initial 
configuration must be \fbox{00} \fbox{2}; its probability, in the 
two-site approximation, is (00)(02)/(0), where (02)/(0) represents 
the conditional probability for a NN pair in state 02, given one site 
in state 0.  To realize the transition, exactly 
one particle must be transferred from the toppling site to its 
neighbor in the 00 pair; this occurs with probability 
$(2d\!-\!1)/2d^2$, as before.  The rate for this
process is then given by
\[
\frac{(2d\!-\!1)^2}{2d^2} \frac{(00)(02)}{(0)}
\]
where the additional factor of $2d\!-\!1$ represents the number of 
possible locations for the neighbor in state 2. (Note that in the 
{\it loss} term for (00) this rate is multiplied by 2 to account 
for the mirror-symmetric process.)
Proceeding in this manner we obtain the rates for each of the 17 
allowed transitions. These are used to generate the equations of 
motion for the pair probabilities, which are then integrated
using a fourth-order Runge-Kutta scheme.

We find $\zeta_c = 0.75$ in 1-d (just as for the unrestricted 
model), and $\zeta_c = 0.63$ in 2-d.  
(The corresponding simulation values are
0.92965 and 0.71127, respectively, as discussed in the following section.)  
The cluster approximation predictions for the active-site density 
are compared with simulation results in Fig. 3.
An interesting qualitative result of the 2-site
approximation is that active sites are {\it anticorrelated}, i.e.,
$(22) < (2)^2$.  This is expected on physical grounds, since, to
become active, a site must have a NN that has toppled recently.

\subsection{Cooperative rule}

For the cooperative rule, the 
evolution equations for the probabilities $(0)$, $(1)$, and $(2)$ are
\begin{equation}
\frac {d}{dt}(0)=-\frac {1}{2} (020)+(121),
\end{equation}
\begin{equation}
\frac {d}{dt}(1)= (020)-2 (121)  ,
\end{equation}
and
\begin{equation}
\frac {d}{dt}(2)=-\frac {1}{2} (020)+(121) .
\end{equation}
To obtain the 1-site approximation we factorize all joint probabilities.
There is then only one independent equation, for example,
\begin{equation}
\frac {d}{dt}(2)=-(2) \left[ \frac {1}{2} (0)^2 +(1)^2 \right].
\end{equation}
In the stationary state this gives $(0)=\sqrt{2} (1)$ from
which it follows that
\begin{equation}
(2)=\frac{3+\sqrt{2}}{7} \left[ \zeta - (\sqrt{2}-1) \right] .
\end{equation}
The critical density is then $\zeta_c = \sqrt{2} - 1 \simeq 0.41421$.

The smaller value of $\zeta_c$ here, as compared with the
independent rule, reflects the fact the cooperative rule tends to
maximize the number of active sites generated.
We show below that the critical density $\zeta_c$ of the independent model
is in fact slightly {\it lower} than that of the cooperative one.
The reason for this is not immediately apparent from the transition rates,
but would appear to lie in subtle
correlations induced by the dynamics, that are not evident at the
1-site level.

\section {Simulation Results}

\subsection{Independent rule}

We performed extensive simulations of the height-restricted
FES with independent
toppling rule in one and two dimensions.
The initial condition is generated by
distributing $\zeta L^d$ particles randomly among the $L^d$ sites,
avoiding occupancy of any site by more than two particles.  This
yields an initial distribution that is spatially homogeneous,
and uncorrelated. The dynamics begins once all
the particles have been placed on the lattice.  The particle
number is, of course, conserved by the dynamics.

In one dimension we study system sizes ranging from
$L = 100$ to $5000$ sites; in two dimensions the
system comprises $L \times L$ sites with $L = 10$, 20, 40,...,320.
For each $L$ we study a range of particle densities
$\zeta \equiv N/L^d$. 
The simulations consist of $N_s$ independent runs, extending to a
maximum time $t_m$.  
(In one dimension, for example, we used $N_s=10^5$, 
$t_m=4000$ for $L=100$, and  $N_s= 2000$, $t_m =2 \times 10^6$, for
$L=5000$.  In two dimensions these parameters varied from 
$N_s = 10^5$, $t_m=1000$, for $L=10$, to $N_s=2 \times 10^4$ and
$t_m = 8 \times 10^4$ for $L=320$.)

Our first task is to locate the critical density
$\zeta_c$; to this end we study the active-site density $\rho_a(t)$,
its second moment $\overline{\rho_a^2}(t)$, and the survival probability $P(t)$.
The second moment is used to evaluate the
ratio $m(t) \equiv \overline{\rho_a^2}(t)/\rho_a^2(t)$.
Figures 4 and 5 show typical results for $\rho_a (t)$
and $P(t)$, respectively.  
$\rho_a (t)$ relaxes to a well-defined 
stationary value, $\rho_a(\zeta,L)$, (similarly for $m$),  
while the exponential decay
of of $P(t)$ allows one to extract an associated 
lifetime, $\tau(\zeta,L)$.  The stationary values,
$\rho_a(\zeta,L)$ and $m(\zeta,L)$ are obtained by discarding the initial,
transient portion of the data, and performing averages over the remainder,
weighted by $P(t)$, which measures the effective sample size.

In a fixed-energy sandpile  
of linear extent $L$, we can 
only vary $\zeta$ in increments of $1/L^d$. To circumvent
this limitation,
work we adopt a strategy employed in a recent
study of the pair contact process \cite{pcpcmp}.  Given simulation
results for the stationary values of $\rho_a$ and $m$, and of the
survival time $\tau$, 
for a certain system size, we form least-squares cubic fits to these data,
permitting interpolation to arbitrary $\zeta$ values within the interval
studied.  Thus, for each $L$, we regard $\rho_a$, $m$, and $\tau$ as a 
functions of a {\it continuous} variable $\zeta$.  
(Since the properties of a finite system are nonsingular, 
the interpolation procedure seems quite natural.)
Data sets for $m$,
and associated cubic fits, are shown in Fig. 6.

A well known criterion for criticality is size-independence of
order-parameter moment ratios, typically in the form of ``crossings" of
Binder's reduced fourth cumulant \cite{binder}.  Moment-ratio crossings
have also proven useful for fixing the critical parameter value at
absorbing-state phase transitions 
\cite{pcpcmp,rdjaff,jaffrd}.
We determine the value
$\zeta_{cr}(L,L')$ for which $m(\zeta,L) = m(\zeta,L')$, for
successive $L$ values. Extrapolating these data to $L \to \infty$ yields 
our estimate for $\zeta_c$;  
Fig. 7 illustrates the procedure.
Evidently the crossing values $\zeta_{cr}(L,L')$ converge quite rapidly.
In two dimensions, the crossings are well described by the form
$\zeta_{cr}(L,L') \simeq \zeta_c + aL^{-b}$ where $a$ is an amplitude and
$b \simeq 2.72$.  

Analysis of the moment-ratio crossings yields
$\zeta_c = 0.92965(3)$ in 1-d and $\zeta_c = 0.711270(3)$ in 2-d,
where the figures in parentheses denote uncertainties.  For comparison,
we note the values for the {\it unrestricted} version of the model:
0.94885(7) in 1-d, 0.71695(5) in 2-d.  Thus the height restriction yields 
a rather small shift in $\zeta_c$, by about 2\% in one dimension, and 
0.8\% in 2-d.
This is reasonable since, in the unrestricted model 
(near its critical point), only a small
fraction of the sites have $z > 2$.
The critical values of the moment ratio are: $m_c = 1.1596(4)$ in
1-d, and 1.347(2) in 2-d.  While these differ significantly from
the corresponding values for the directed percolation (DP) universailty
class [1.1735(5) and 1.3257(5) in 1-d and 2-d, respectively \cite{rdjaff}],
the moment ratios for the two classes are very similar.

In studies of absorbing-state phase transitions \cite{marro}, including
fixed-energy sandpiles \cite{dvz,m1d}, it is
common to determine the critical point by seeking a power-law dependence of
the order parameter ($\rho_a$ in the present instance) and the relaxation
time on the system size $L$.  The former is governed by 

\begin{equation}
\rho_a (\zeta,L) = 
L^{-\beta/\nu_{\perp}} {\cal R} 
(L^{1/\nu_{\perp}} \Delta) \;,
\label{actfss}
\end{equation}
as expected on the basis of finite-size scaling \cite{fss}.
(Here $\Delta \equiv \zeta - \zeta_c$, and
${\cal R}$ is a scaling function.)  Thus at the critical point 
($\Delta = 0$) we expect $\rho_a (\zeta_c,L) \sim 
L^{-\beta/\nu_{\perp}}$; for the lifetime
one has $\tau (\zeta_c,L) \sim L^{\nu_{||}/\nu_{\perp}}$.

With $\zeta_c$ in hand, we may verify the power-law dependence of
the order parameter and the lifetime on system size, as in 
Eq. (\ref{actfss}), by interpolating the simulation data to the critical
value $\zeta_c$.  Figure 8 shows that $\rho_a$ indeed has a power-law 
dependence on $L$; a similar plot (not shown) yields the same 
conclusion for $\tau$.  From the data for the four 
largest systems, we then obtain (via least-squares linear fits),
the exponent ratios $\beta/\nu_\perp$ and $\nu_{||}/\nu_\perp$
listed in Table II.  (The uncertainties reflect two contributions: one due to 
the uncertainty of the fit, the other, dominant one, due to the uncertainties
in the values of $\rho_a$ and $\tau$ for each $L$.  The latter includes the
effects of uncertainty in $\zeta_c$.)

To determine the exponent $\beta$ we analyze the results for $\rho_a$
in the portion of the supercritical regime where the graph of $\ln \rho$
versus $\ln \Delta$ follows a power law.  
In two dimensions this procedure yields 
$\beta=0.661(3)$, 0.661(2), 0.654(3), and $0.655(2)$ for $L=20$, 40, 80, 
and 160, respectively, leading to an estimate of $\beta=0.656(5)$.
Fig. 9, a scaling plot of 
$L^{\beta/\nu_{\perp}} \rho_a (\zeta,L)$ versus 
$ L^{1/\nu_{\perp}} \Delta$ for various system sizes,
shows a good data collapse, verifying the finite-size scaling
hypothesis for the order parameter, and yielding $\nu_\perp = 0.85$. 

In one dimension it turns out that no power laws are seen if we use
$\zeta_c = 0.92965$ as determined from the FSS analysis described above.
Quite clean power law dependence of $\rho_a$ is observed, however, if we
use an $L$-dependent {\it effective} critical point $\zeta_{c,L}$ in 
the analysis.  We determine $\zeta_{c,L}$ by optimizing the linearity
of $\ln \rho_a$ as a function of $\ln \Delta$, and maximizing the number
of data points that may reasonably be fit by the power law.  (For 
$L=1000$ for example, we are able to fit 15 points with a 
correlation coefficient of $0.99996$.)  The resulting values of
$\zeta_{c,L}$ and $\beta$ are listed in Table III.  
Extrapolating the effective critical densities to infinite $L$
(via linear regression versus $L^{-1/\nu_\perp}$) yields
$\zeta_c = 0.9298(4)$, consistent
with our our estimate based on moment-ratio crossings.
A similar extrapolation gives $\beta = 0.412(4)$
We determine the exponent $\nu_\perp$ via a data-collapse 
analysis, as in the 2d case (see Fig. 10).  
We obtain a good data collapse for 
$1/\nu_\perp$ in the range 0.60 to 0.62, leading to the estimate
$\nu_\perp = 1.64(4)$.  

\subsection{Cooperative rule}

We performed extensive simulations of the cooperative
model in one dimension with system sizes again ranging from $L=100$
to $L=5000$. 
In this case we prevented the system from falling into an 
absorbing configuration 
by maintaing at least two active sites. 
(If there are only two active sites,
transitions which decreases the number of active sites are not
permitted. Actually, there is only one transition of this sort,
$020\rightarrow 101$.)
The density of active sites $\rho$ is then
always $\geq 2/L$. 
But since the stationary value of $\rho_a$ at the critical
point is $\sim L^{-\beta/\nu_\perp}$, with 
$\beta/\nu_\perp \simeq 1/4$,
this should have a minimal effect on critical properties.
Figure 11 shows the stationary active-site
density as a function of the particle density
for several values of $L$.

We first analyzed
the stationary critical properties of the model by
means of the finite size scaling relation, Eq. (\ref{actfss}). 
The critical density was obtained by plotting
$\rho_a$ versus $L$ for several $\zeta$ values,
as shown in Fig. 12.
(As before, values of $\rho_a$ for densities between those accessible
for a given $L$ were obtained via interpolation.)
Using the criterion of power law dependence
of the order parameter on system size,
we find $\zeta_c=0.9788(1)$ and
$\beta/\nu_{\perp}=0.245(5)$. 
As an alternative determination of
$\zeta_c$ we used moment-ratio crossings. Figure 13
shows the moment ratio $m$ as a function of $\zeta$ for
$L=2000$ and $L=5000$. The two curves cross at
$\zeta_c=0.9788$, confirming the previous result.

Having obtained the critical particle density,
we used it to find the critical exponent $\beta$ governing the
order parameter. Figure 14 is a log-log plot of
$\rho_a$ versus $\zeta-\zeta_c$ for several values of $L$.
The slope of the straight line fitted to the data points
for $L=5000$ gives $\beta=0.417(1)$. From this,
and our previous result for $\beta/\nu_{\perp}$, we
obtain $\nu_{\perp}= 1.70(4)$.

We also performed time-dependent simulations at the critical
density, to measure the growth of the number of active sites.
Here, each trial began
with just one active site. For a given particle
density $\zeta$, this was realized
by placing a particle at each of $\zeta L -1$ distinct sites, chosen
at random. One of these sites was then selected randomly, and
another particle placed there, rendering it
active. In a lattice of size $L=10000$, we performed
from $5000$ to $6000$ trials of this kind, to determine the 
mean number of particles, $n(t)$, averaged over all trials
(including those that fall into an absorbing configuration 
prior to time $t$).
At the critical point, and for a sufficient large system, $n(t)$ is
expected to increase asymptotically as a power law
\begin{equation}
n(t) \sim t^\eta
\end{equation}
where the exponent $\eta$ is related to the exponent
$z=\nu_{||}/\nu_{\perp}$
by the scaling relation \cite{mdvz,torre},
\begin{equation}
z=\frac {1}{\eta} \left( d-2\frac {\beta} {\nu_{\perp}} \right),
\end{equation}
in $d$ dimensions.
Our data for $n(t)$ at $\zeta_c = 0.9788$ do in fact 
follow a power law, and yield 
the estimate $\eta=0.330(5)$.
Using our previous result for $\beta/\nu_{\perp} $ we 
then obtain $z=1.54(5)$.

\section{Discussion}

We studied the scaling behavior of fixed-energy sandpiles that
follow a stochastic dynamics similar to that of the Manna model,
but with a height restriction $z_i \leq 2$.  
Both versions of the model (i.e., the independent and cooperative
toppling rules), exhibit a continuous phase 
transition between an absorbing 
state and an active one at a critical particle density
$\zeta_c$.  
One- and two-site cluster approximations 
do not yield very accurate predictions for the critical density
(as is to be expected), but they correctly predict the
continuous nature of the transition.  

As shown in Table II, the critical exponents for the present
models appear to be the same as for the unrestricted case.
In fact, there is excellent agreement between the exponent values
for the restricted and unrestricted models, except for the
exponent $z = \nu_{||}/\nu_\perp$ in one dimension.  As was noted
in Ref. \cite{m1d}, however, obtaining a reliable estimate for
the exponent $z$ from simulations is quite difficult in one dimension.
(In two dimensions the estimates for $z$ are in excellent agreement.
It appears that the relaxation dynamics is anomalous in one
dimension, as suggested in \cite{m1d}.)

Studies of a reaction model suggest a common
universality class, distinct from that of directed percolation, 
for absorbing-state phase transitions in  
which the order parameter is coupled to a second field that relaxes
diffusively in the presence of activity \cite{ale00}.
Manna's stochastic sandpile falls in this category, with the local
particle density $\zeta({\bf x},t)$ playing the role of the
second field.
Our results show that height restrictions and perturbations 
of the toppling rule do not alter the critical exponents, 
if they preserve the above-mentioned features,
supporting universality in critical behavior far from equilibrium.
\vspace{2em}

\noindent{\bf Acknowledgements}
\vspace{1em}

We thank Paulo Alfredo Gon\c{c}alves Penido for helpful comments.
This work was supported by CNPq and CAPES, Brazil.
\vspace{2em}

\noindent$^\dagger$email: dickman@fisica.ufmg.br

\newpage
\begin{table}
\begin{center}
\begin{tabular}{|r|c|c|}
Transition         &  \multicolumn{2}{c|}{ Probability}    \\        
\hline
		   & Independent & Cooperative  \\ 
\hline
$020 \to 101$      &    1/2      &     1/2      \\ 
$\;\;\;\; \to 200$ &    1/4      &     1/4      \\ 
$\;\;\;\; \to 002$ &    1/4      &     1/4      \\ 
\hline  
$120 \to 201$      &    1/2      &     1/2      \\ 
$\;\;\;\; \to 102$ &    1/4      &     1/2      \\ 
$\;\;\;\; \to 210$ &    1/4      &      0       \\ 
\hline  
$220 \to 202$      &    1/4      &      1       \\
$\;\;\;\; \to 211$ &    1/2      &      0       \\
$\;\;\;\; \to 220$ &    1/4      &      0       \\
\hline
$121 \to 202$      &    1/2      &      1       \\
$\;\;\;\; \to 112$ &    1/4      &      0       \\
$\;\;\;\; \to 211$ &    1/4      &      0       \\
\hline
$122 \to 212$      &    3/4      &      1       \\ 
$\;\;\;\; \to 122$ &    1/4      &      0       \\ 
\end{tabular}
\end{center}
\label{tab1}
\noindent{Table I. Transition probabilities for the independent and cooperative
toppling rules in one dimension.  The transition probabilities are
symmetric under reflection.}  
\end{table}

\begin{table}
\begin{center}
\begin{tabular}{|c|c|c|c|c|}
Model  &  $\zeta_c$ & $\beta/\nu_\perp$ & $\nu_{||}/\nu_\perp$ & $\beta$     \\        
\hline
\hline
Indep. 1d & 0.92965(3) & 0.247(2)  & 1.45(3) & 0.412(4)\\ 
Coop. 1d  & 0.9788(1)  & 0.245(5)  & 1.54(5) & 0.417(1) \\ 
Unrst. 1d & 0.94885(7) & 0.239(11) & 1.66(7) & 0.42(2) \\ 
\hline  
Indep. 2d & 0.711270(3)& 0.774(3) & 1.572(7) & 0.656(5) \\ 
Unrst. 2d & 0.71695(5) & 0.78(2)  & 1.57(4)  & 0.64(1)  \\ 
\end{tabular}
\end{center}
\label{tab2}
\noindent{Table II. Critical parameters of restricted and unrestricted sandpiles.  
Figures in parentheses denote uncertainties.  Results for the unrestricted
models are from Refs. \cite{m1d} (1-d) and \cite{fes2} (2-d).
}  
\end{table}

\begin{table}
\begin{center}
\begin{tabular}{|c|c|c|}
$L$  &  $\zeta_{c,L}$ & $\beta_L$   \\        
\hline
\hline
500   & 0.9256(1)  & 0.465(3)  \\ 
1000  & 0.9273(1)  & 0.441(4)  \\ 
2000  & 0.92815(5) & 0.431(4)  \\ 
2000  & 0.92845(5) & 0.423(4)  \\ 
\hline  
$\infty$ & 0.9298(4) & 0.412(4)   \\ 
\end{tabular}
\end{center}
\label{tab3}
\noindent{Table III. Effective size-dependent critical density and apparent  
exponent $\beta_L$, and $L \to \infty$ extrapolated values, for
the 1-d model with independent toppling rule.
}  
\end{table}

\newpage
\noindent FIGURE CAPTIONS
\vspace{1em}

\noindent FIG. 1. Transitions between states of a single site. 
`$\times$' denotes a
forbidden transition; diagonal entries are irrelevant.
\vspace{1em} 

\noindent FIG. 2. Transitions between configurations of a NN pair 
of sites; `$\times$' denotes a
forbidden transition; diagonal entries are irrelevant.
\vspace{1em} 

\noindent FIG. 3. 
The stationary active-site density in the two-dimensional
restricted-height sandpile as predicted by the one-site
and two-site approximations, compared with the simulation result.
\vspace{1em} 

\noindent FIG. 4. 
Evolution of the active-site density in the one-dimensional 
sandpile with height restriction (independent rule).  
$L\!=\!1000$; $\zeta\!=\!0.93$.
\vspace{1em}

\noindent FIG. 5.  
Evolution of the survival probability $P(t)$, for the same
parameters as in Fig. 4. 
\vspace{1em}

\noindent FIG. 6. 
Stationary moment ratio, $m(\zeta)$, in 
the one-dimensional model (independent rule).
Squares: $L\!=\!500$; $+$: $L\!=\!2000$.  Curves are cubic fits to the
data.
\vspace{1em}

\noindent FIG. 7.  
Moment-ratio crossing values $\zeta_{cr}$ versus reciprocal system
size in 1-d (independent rule).
\vspace{1em}

\noindent FIG. 8. 
Stationary active-site density $\rho_a$ versus system size $L$ at
the critical point in 1-d ($+$) and 2-d (squares) 
(independent rule).
\vspace{1em}

\noindent FIG. 9. 
Scaling plot of the stationary active-site density in the 2-d
(independent rule).  System sizes $L\!=\!20$ (filled squares),
40 (open squares), 80 ($\times$) and 160 ($+$).
\vspace{1em}

\noindent FIG. 10. 
Scaling plot of the stationary active-site density in 1-d
(independent rule).  System sizes $L\!=\!500$ ($\circ$),
1000 ($\diamond$), 2000 ($\times$) and 5000 ($+$).
\vspace{1em}

\noindent FIG. 11. 
Stationary active-site density in 1-d versus particle density,
for various system sizes (cooperative rule).  
\vspace{1em}

\noindent FIG. 12. 
Stationary active-site density versus system size in 1-d  
(cooperative rule).
\vspace{1em}

\noindent FIG. 13. 
Stationary moment ratio in
the one-dimensional model (cooperative rule).
\vspace{1em}

\noindent FIG. 14. 
Stationary active-site density 
versus $\zeta \!-\! \zeta_c$ for various system sizes 
(cooperative rule, 1-d).  

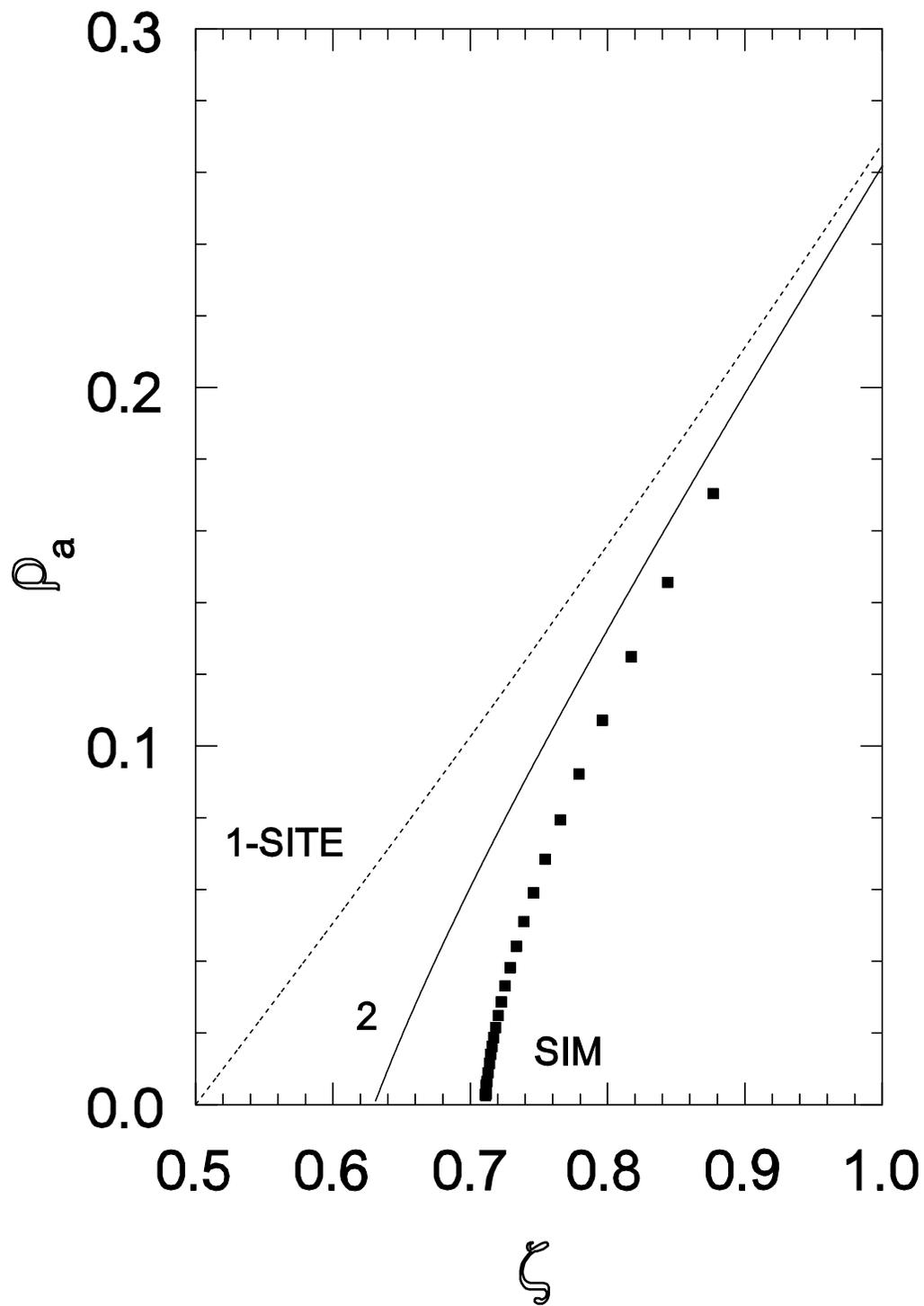
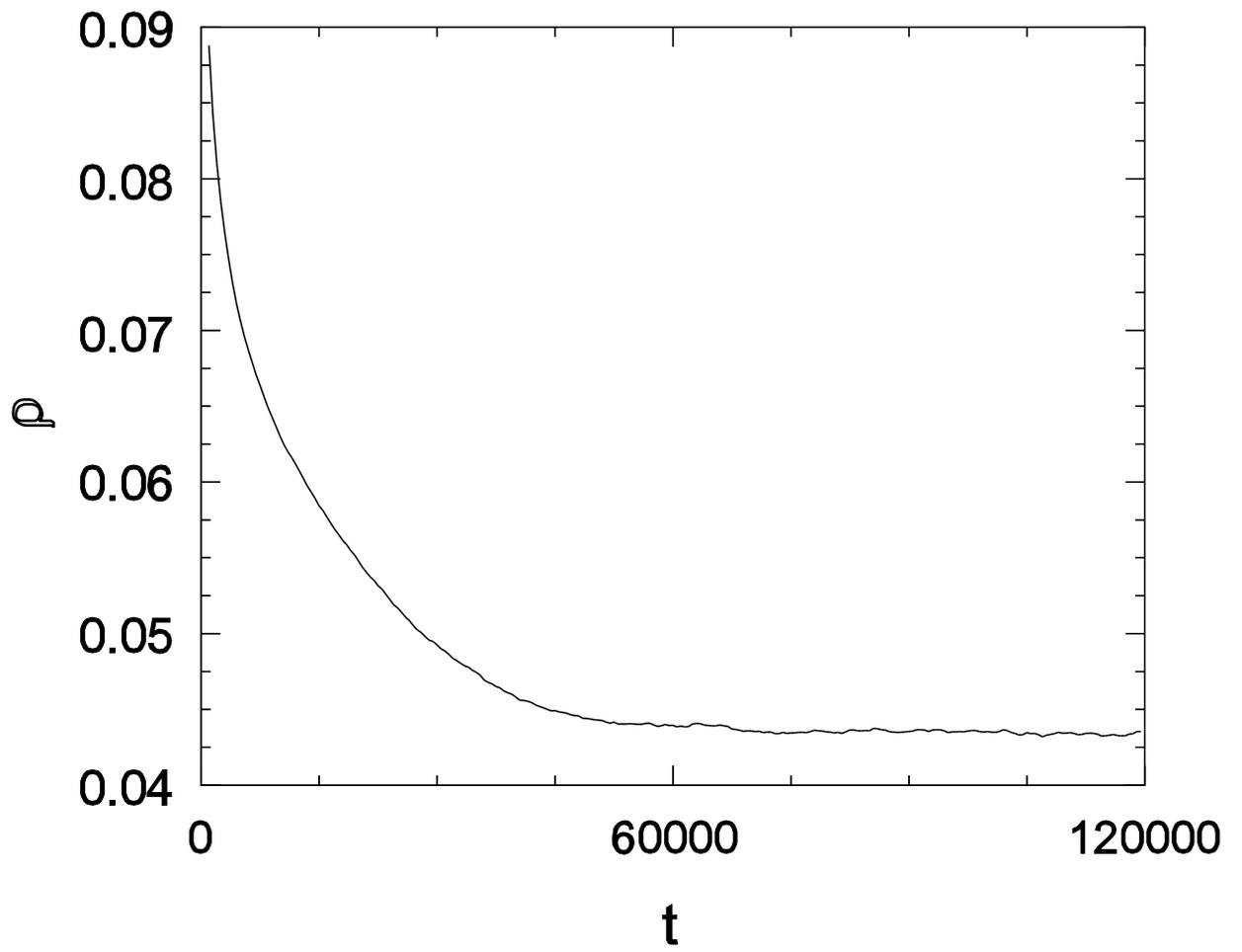
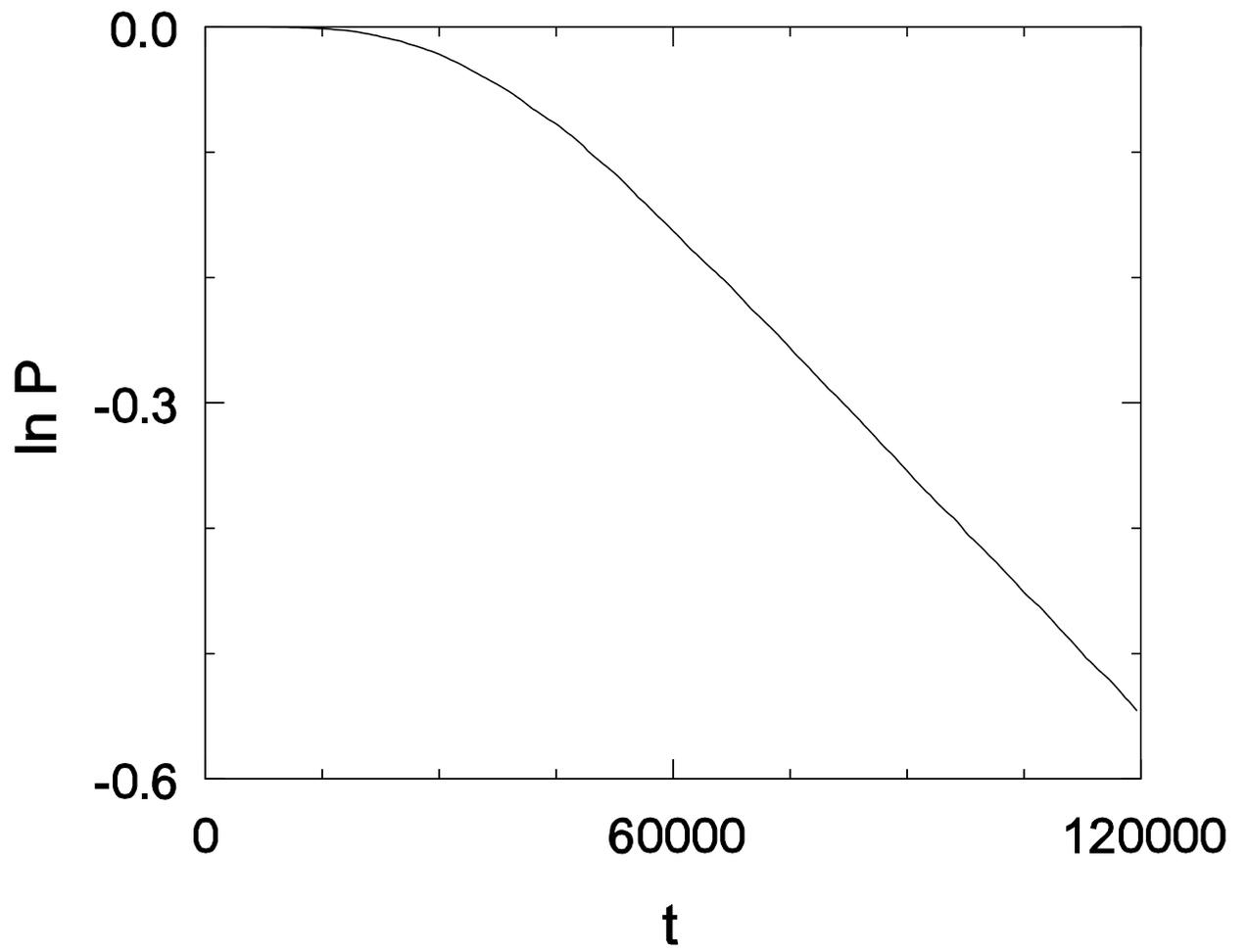
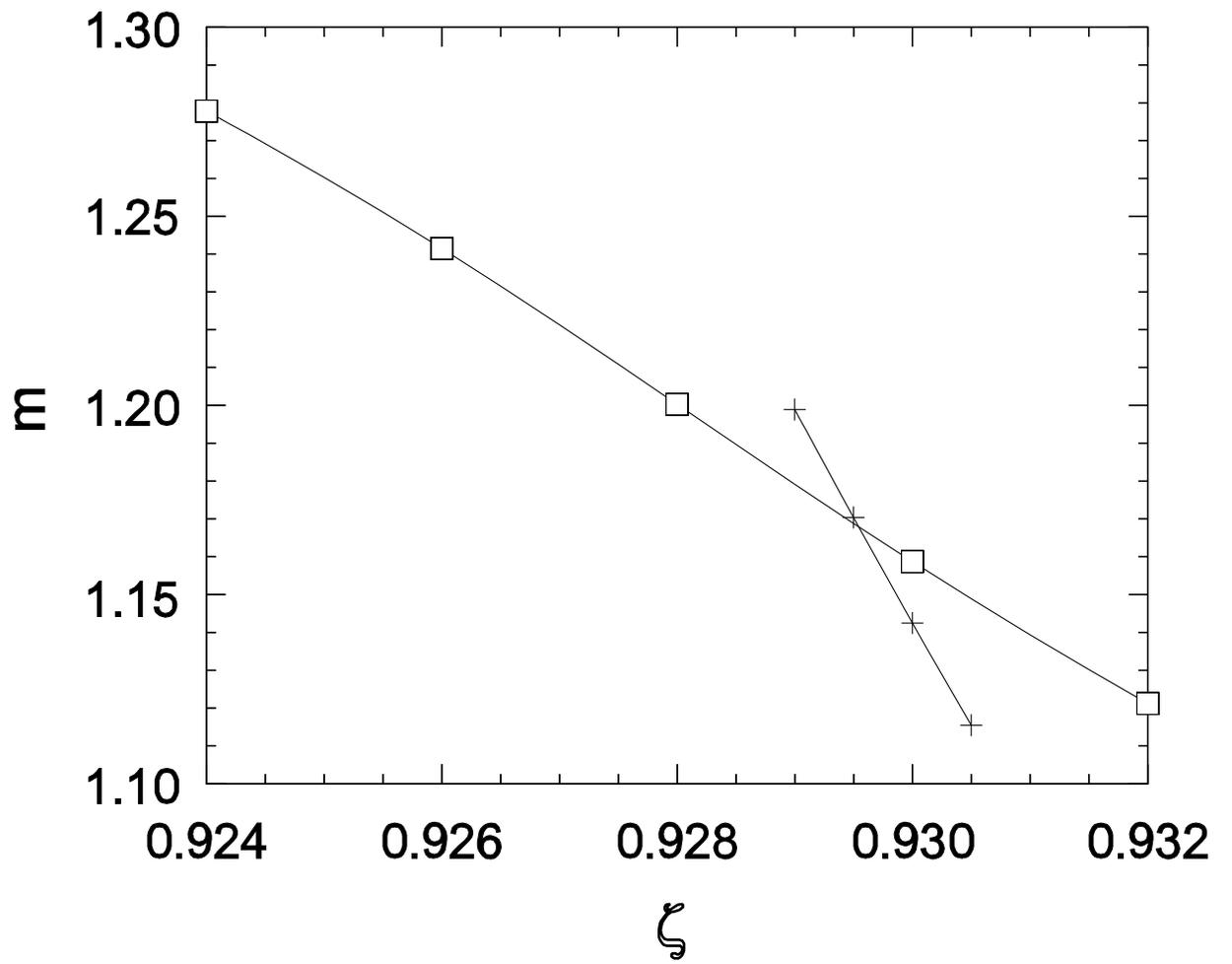
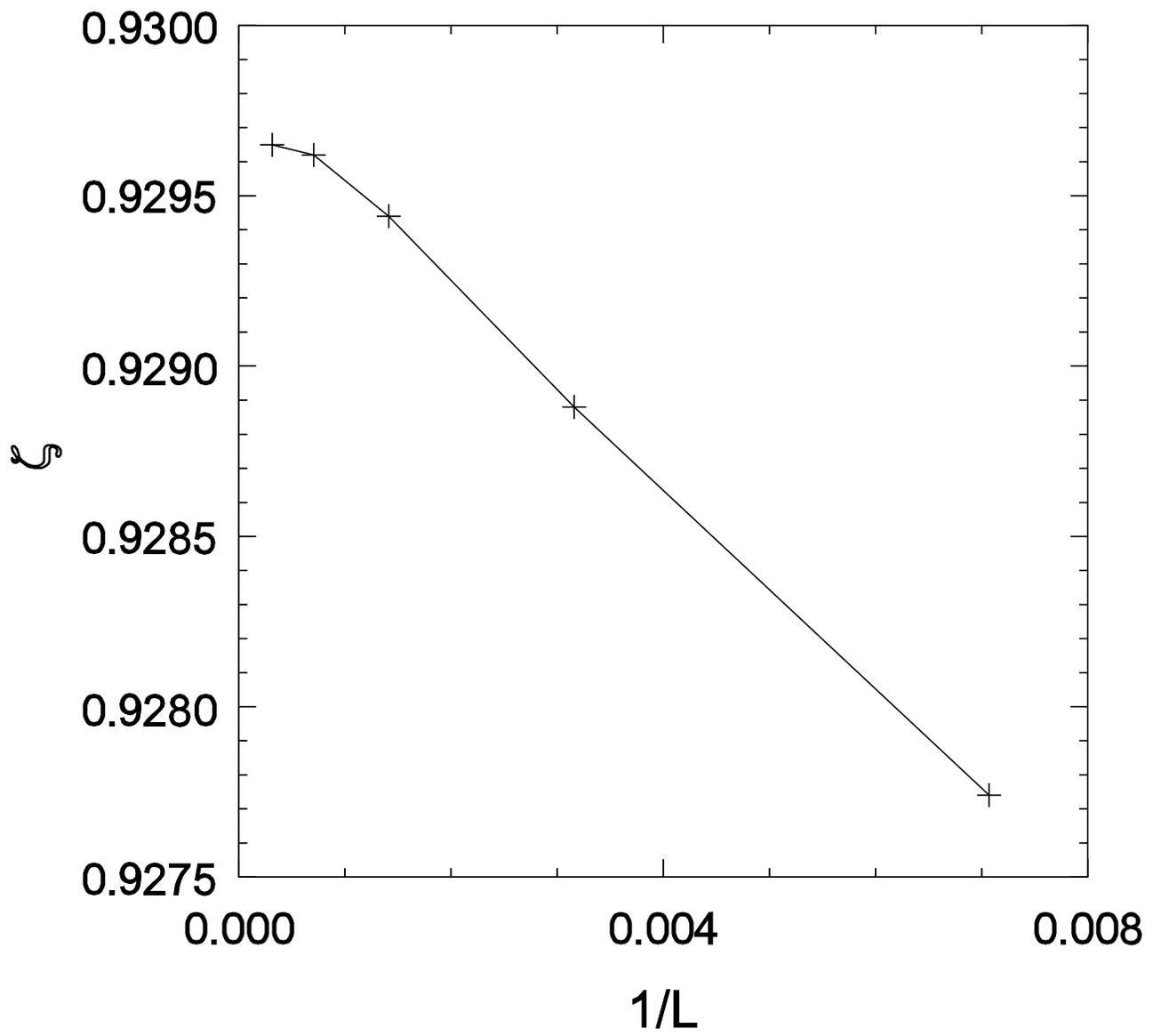
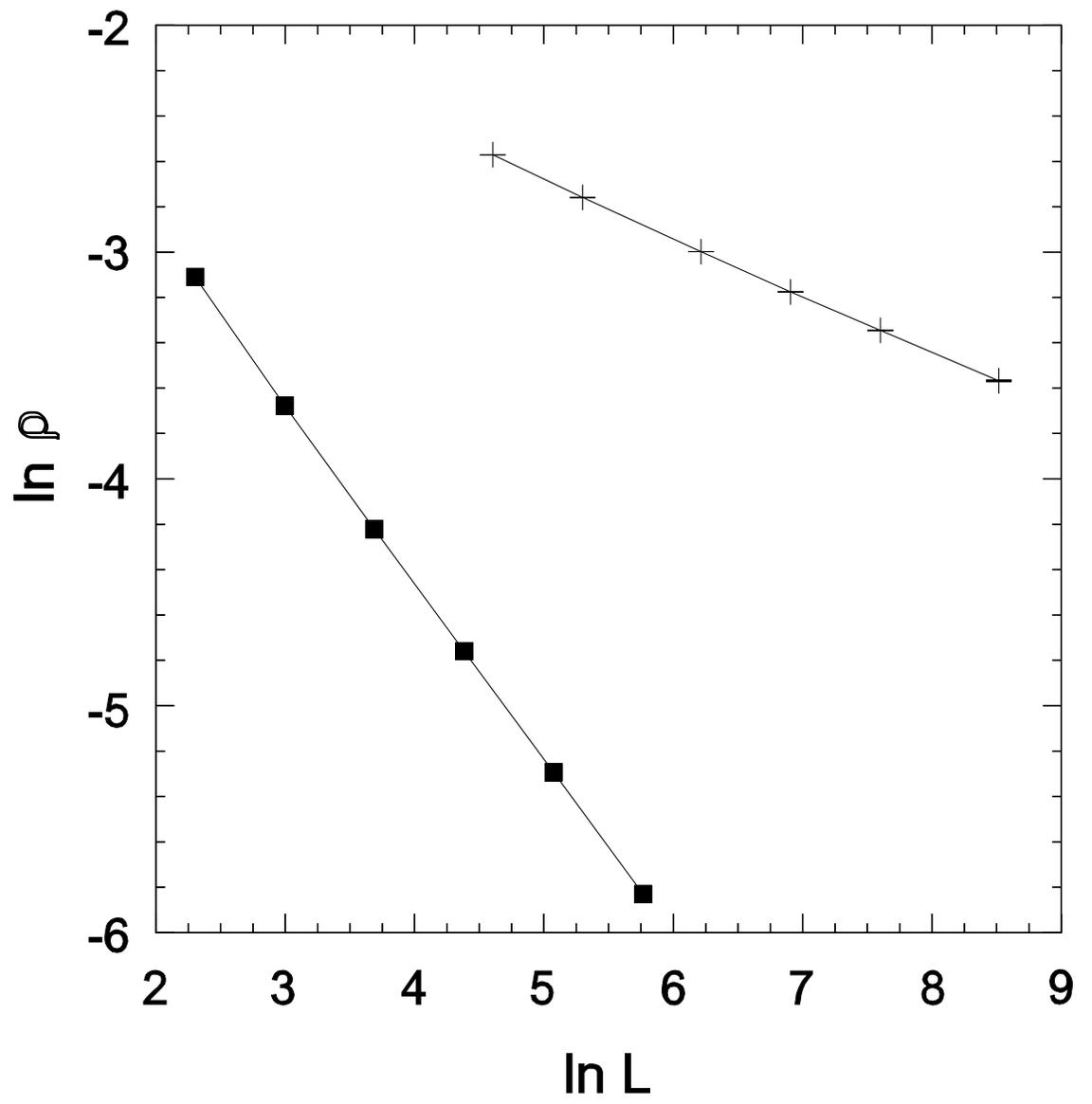
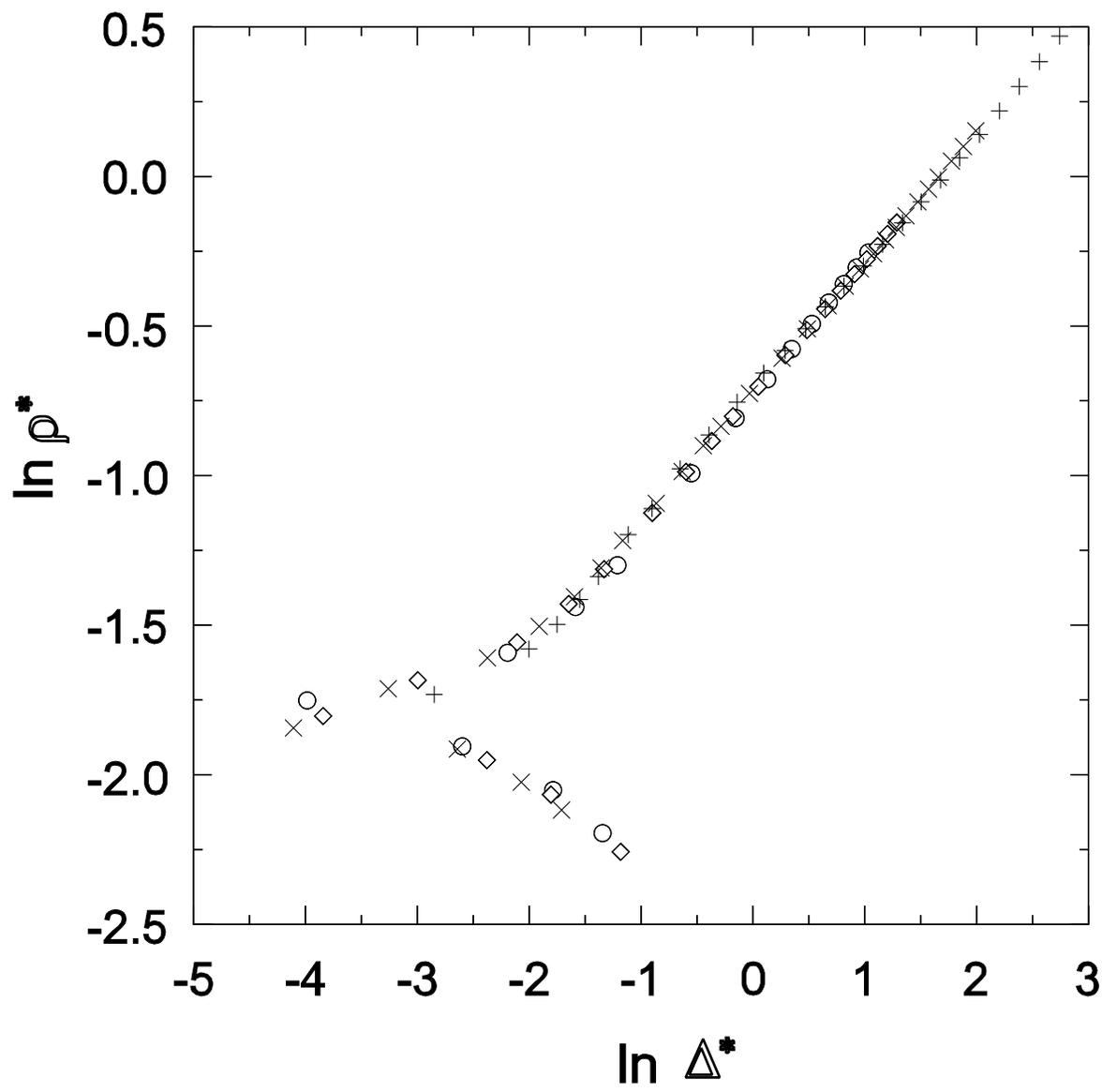
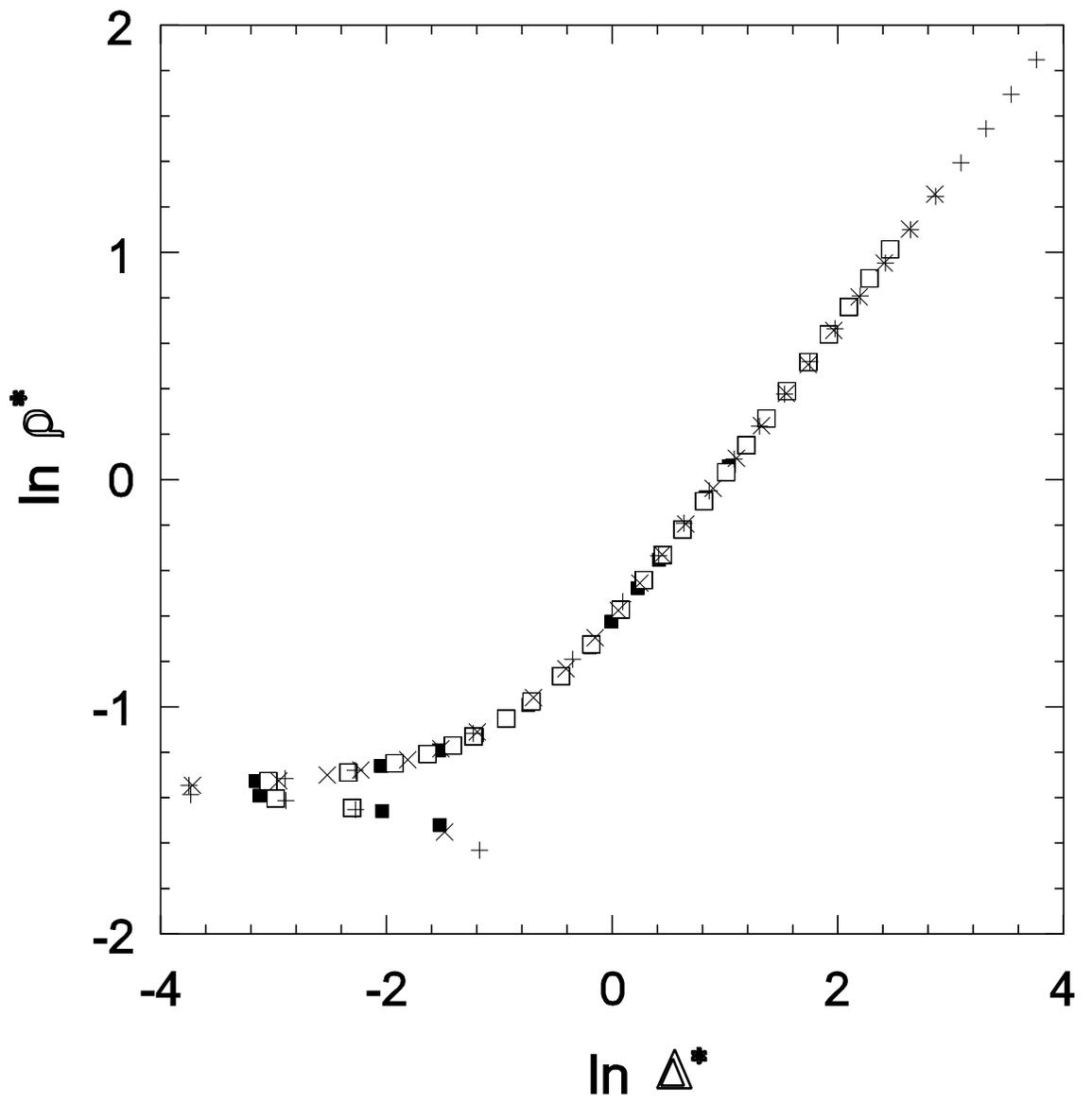
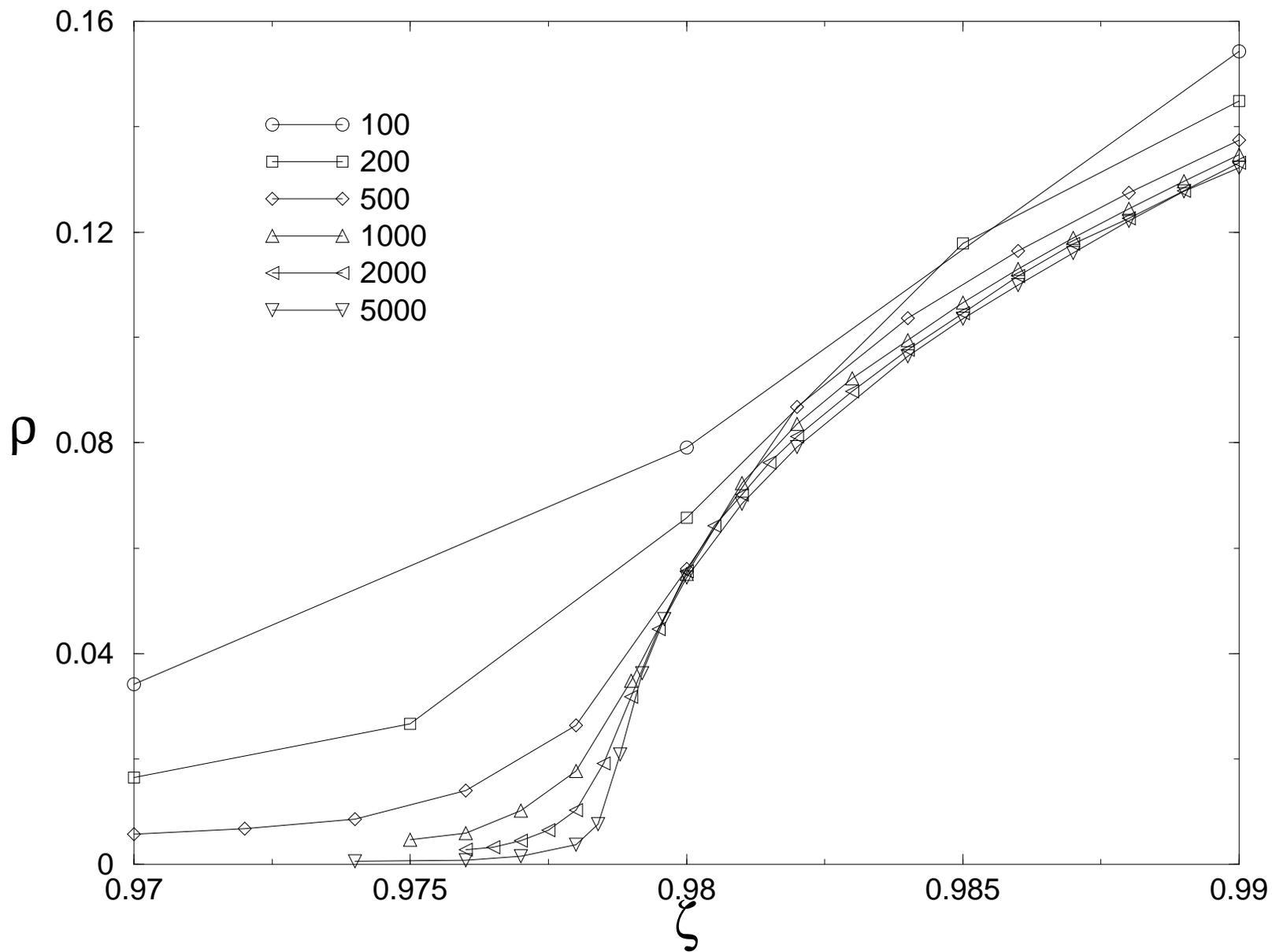
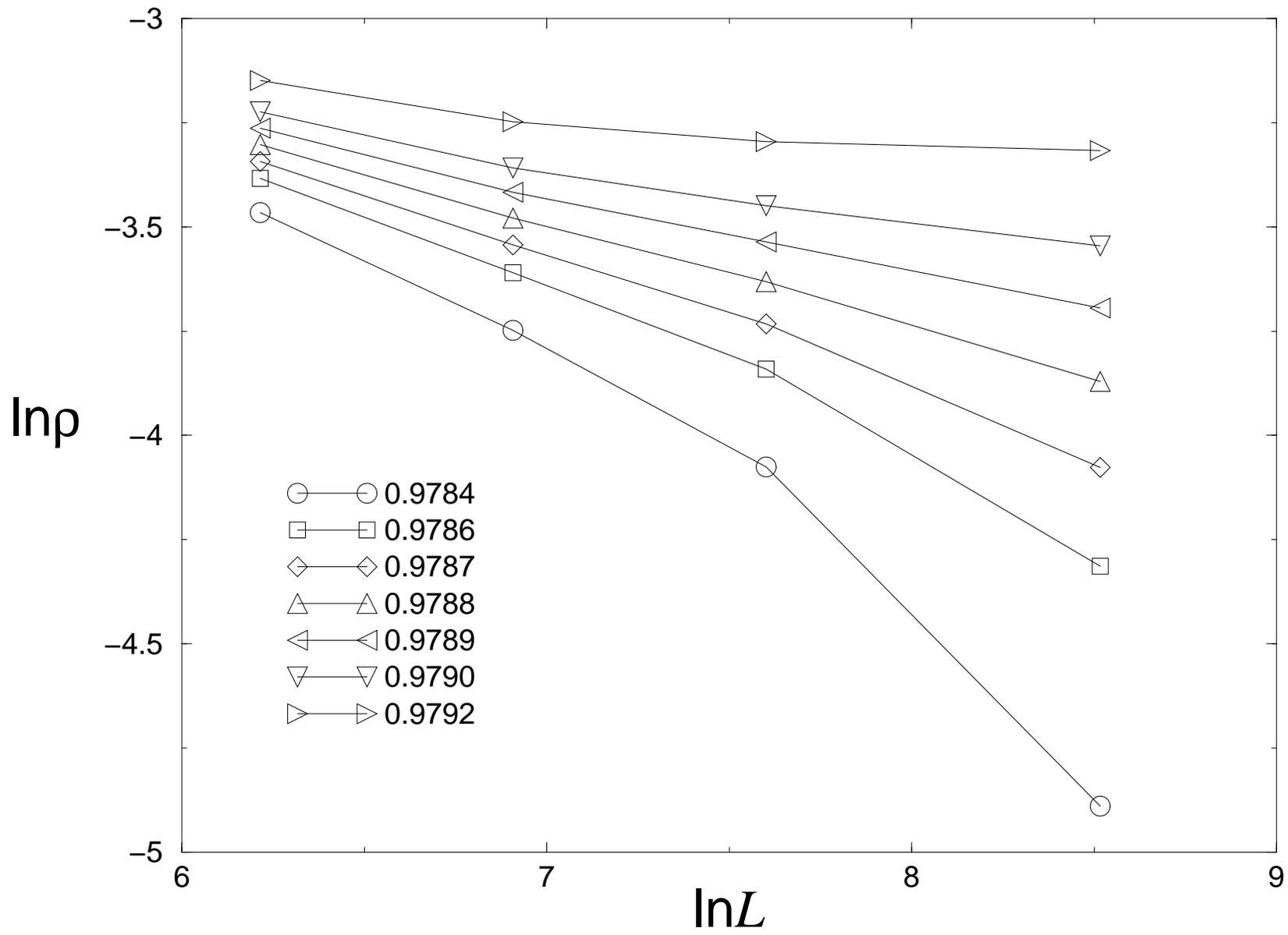
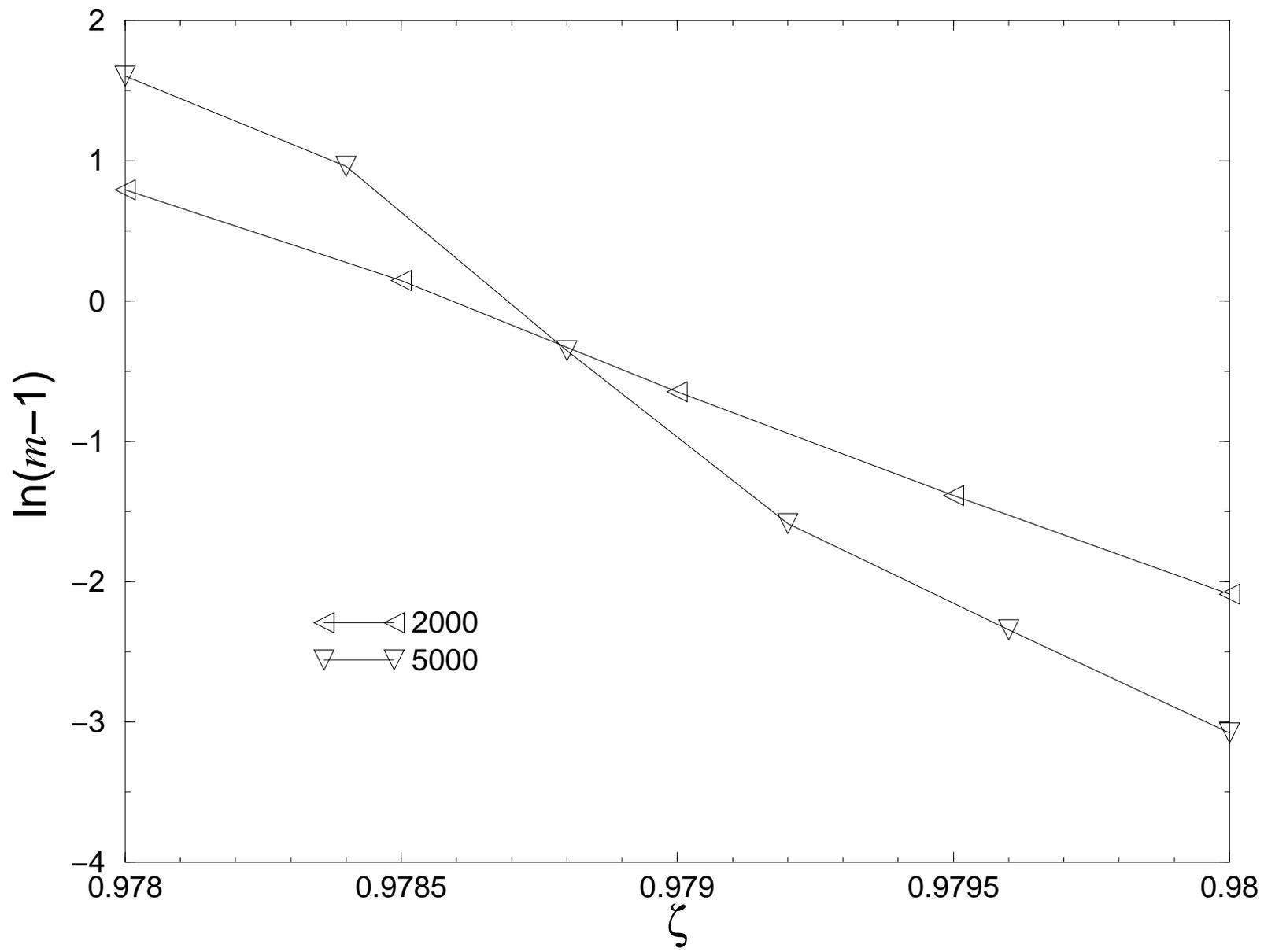
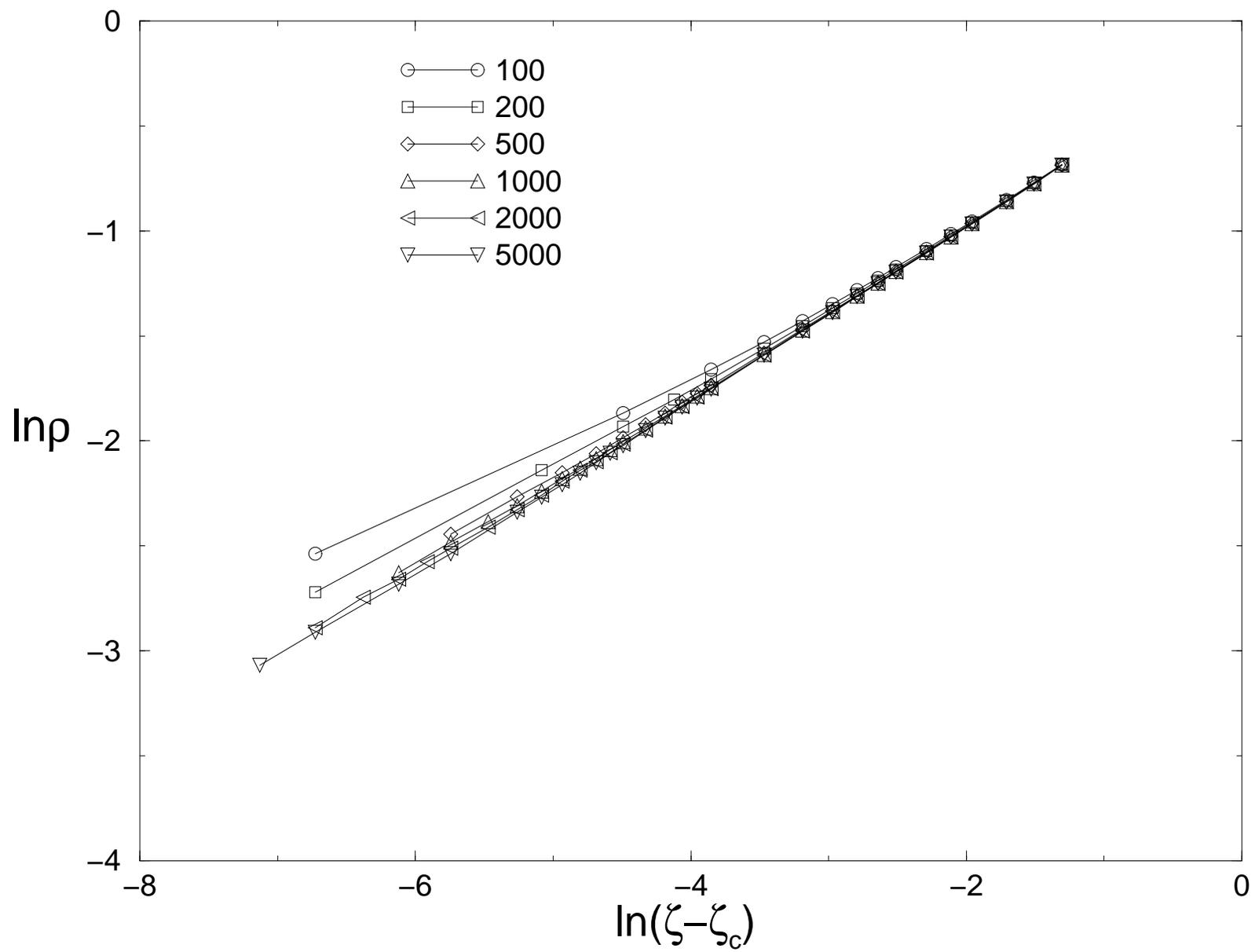
\end{document}